\documentclass[twocolumn,pra,superscriptaddress]{revtex4-1}
\usepackage{palatino}
\usepackage[latin1]{inputenc}
\usepackage{epsf}
\usepackage{amsmath,amssymb}
\usepackage{latexsym}
\usepackage{calc}
\usepackage{color}
\usepackage{shadow}
\usepackage{epsfig}
\usepackage[usenames,dvipsnames]{xcolor}
\usepackage{changes}
\def\bea{\begin{eqnarray}}
\def\eea{\end{eqnarray}}
\def\ben{\begin{equation}}
\def\een{\end{equation}}
\def\benu{\begin{enumerate}}
\def\enu{\end{enumerate}}

\def\dulR{{\underline{\underline{\bf R}}}}
\def\dulr{{\underline{\underline{\bf r}}}}

\def\n{n}

\def\1var{(\bx_1...\bx\N)}

\def\br{{\bf r}}
\def\bR{{\bf R}}

\def\bx{{\br t}}

\def\ext{_{\rm ext}}

\begin{document}
\title{Electronic Non-adiabatic Dynamics in Enhanced Ionization of Isotopologues of H$_2^+$ from the Exact Factorization Perspective}
\author{Elham Khosravi}
\email{elham.etn@gmail.com}
\thanks{Corresponding author}
\address{Nano-Bio  Spectroscopy  Group  and  ETSF,  Universidad  del  Pa\'is  Vasco,
CFM CSIC-UPV/EHU,  20018  San Sebasti\'an, Spain}
\author{Ali Abedi}
\email{aliabedik@gmail.com}
\thanks{Corresponding author}
\address{Nano-Bio  Spectroscopy  Group  and  ETSF,  Universidad  del  Pa\'is  Vasco,
CFM CSIC-UPV/EHU,  20018  San Sebasti\'an, Spain}
\author{Angel Rubio}
\email{angel.rubio@mpsd.mpg.de}
\thanks{Corresponding author}
\address{Max Planck Institute for the Structure and Dynamics of Matter, Luruper Chaussee 149, 22761 Hamburg, Germany}
\address{Nano-Bio  Spectroscopy  Group  and  ETSF,  Universidad  del  Pa\'is  Vasco,
CFM CSIC-UPV/EHU,  20018  San Sebasti\'an, Spain}
\author{Neepa T. Maitra}
\email{nmaitra@hunter.cuny.edu}
\thanks{Corresponding author}
\address{Department of Physics and Astronomy, Hunter College and the Graduate Center of the City University of New York, 695 Park Avenue, New York, New York 10065, USA}
\date{\today}
\pacs{}

\begin{abstract} 
It was recently shown that the exact potential driving the electron's dynamics in enhanced ionization of H$_2^+$ can have large contributions arising from dynamical 
electron-nuclear correlation, going beyond what any electrostatics-based model can provide~\cite{KAM15}. This potential is defined via the exact factorization of the molecular 
wavefunction that allows the construction of a Schr\"odinger equation for the electronic system, in which the potential contains exactly the effect of coupling to 
the nuclear system and any external fields. Here we study enhanced ionization in  isotopologues of H$_2^+$ in order to investigate 
nuclear-mass-dependence of these terms for this process.  We decompose the exact potential into components that naturally arise from the conditional wavefunction, 
and also into components arising from the marginal electronic wavefunction, and compare the performance of propagation on these different components 
as well as approximate potentials based on the quasi-static or Hartree approximation with the exact propagation. 
A quasiclassical analysis is presented to help analyse the structure of different non-electrostatic components to the potential driving the ionizing electron. 
\end{abstract}
\maketitle
\section{Introduction}
The phenomenon of charge-resonance enhanced ionization (CREI), predicted more than twenty years ago~\cite{ZCB93,ZB95,SIC95}, is a prominent example of the complex coupling of electronic motion, ionic motion, and strong laser fields. At a critical range of internuclear separations, the ionization rate of a molecule in a laser field can be orders of magnitude greater than the rate from the constituent atoms. 
The ionization rate has been explained by a quasi-static argument involving instantaneously frozen nuclei in the pioneering 
works of Refs.~\cite{ZB95,CFB98,CB95,SIC95,CCZB96,YZB96,BL12} for which the
time-dependent Schr\"odinger equation (TDSE) is solved for various clamped nuclear (cn) configurations $\dulR_0$, i.e., 
\ben
\label{eq,cn-tdse}
\hat{H}_{el}(\dulr,\dulR_0)\Phi^{cn}_{\dulR_0}(\dulr,t)= i \hbar \partial_t\Phi^{cn}_{\dulR_0}(\dulr,t), 
\een
where 
\ben
\hat{H}_{el}(\dulr,\dulR_0)=\hat{T}_e +\hat{W}_{ee}(\dulr)+\hat{W}_{en}(\dulr,\dulR_0)+\hat{V}_e^{laser}(\dulr,t).
\een
Here $\dulr$ and $\dulR$ are used to collectively denote the
electronic and nuclear coordinates, $\hat{T}_e$ is the electronic kinetic energy operator, and $\hat{W}_{ee}$ is the electron-electron repulsion. 
Furthermore, $\hat{W}_{en}(\dulr,\dulR_0)$ contains the electron-nuclear interaction, which, for a diatomic molecular ion 
with one electron, as will be considered here, has the  form $-Z_1/\vert\br - \bR_1\vert - Z_2/\vert\br - \bR_2\vert$, i.e. a Coulombic double-well. 
The critical internuclear separation for CREI can be qualitatively explained by the following argument: at stretched geometries in a given static field, the energy levels in the
up-field atom Stark-shift upwards while the inner barrier from the internuclear Coulombic potential also grows (Fig.~2 in Ref.~\cite{ZB95}). Provided the field is turned on 
fast enough such that any population in the up-field level (LUMO) does not significantly tunnel back to the down-field atomic level (HOMO), the molecule can rapidly 
ionize over both the inner and outer barriers, which gives rise to the
ionization rate observed to be enhanced by orders of magnitude compared to the atomic rate. 
By requiring that the Stark-shifted LUMO level exceeds the top of both the inner and outer field-modified Coulombic barriers, one finds $R_c = 4.07/I_p$ for the critical
internuclear separation for CREI. The analysis can be generalized to the case of a laser field, where the field's period is shorter than the tunneling time 
\cite{ZB95, ZCB93, TB10, TB11}. 

It has been pointed out that, in actuality, the underlying assumption in this picture of the electron adiabatically following the field is not quite adequate, as the ionization tends to occur in multiple sub-cycle ionization bursts, not at the peak of the field cycles~\cite{TB10,TB11}. 
Moreover,  when applied to an experiment where the molecule's initial geometry is far from where the CREI is expected to happen,
the premise of the quasi-static picture can become a little shaky: in particular, 
the molecule must dissociate to the CREI region, which occurs
predominantly by Coulomb explosion following ionization, but to
actually observe CREI rapidly enough that appreciable
electron density still remains largely un-ionized. In fact, in many experiments, CREI is not observed because too little electron density reaches the CREI region over the course of the applied field~\cite{BSC06,BKPB11,LLDQ03}.
Typically a large fraction of the nuclear density remains near equilibrium, while
a part of it dissociates. So, (i) representing its potential on the electron as a Coulombic double-well is not appropriate, and (ii) the fragment 
velocities can be comparable to the electronic velocities in some dissociating channels questioning the very notion of electrostatic nuclear-electron interactions. In Ref.~\cite{KAM15}, 
it was shown that the nuclear dynamics can contribute significant components to the exact potential that arise from dynamical electron-nuclear correlated motion. Neglect of these contributions leads to severe errors in the prediction of the electronic motion. 

In this paper, we analyse the exact potential in more detail by
providing a complementary decomposition to the one that was studied
in Ref.~\cite{KAM15}, comparing with different 
approximations, studying the mass-dependence, and investigating a
quasiclassical treatment. The rest of the paper is organized as follows:  First, a brief review of
the exact factorization approach is presented in Sec.\ref{sec:exact_fact}, with a focus on the reverse factorization introduced in 
Ref.~\cite{SAMYG14}. This provides us with a TDSE for electrons evolving on a single time-dependent potential that accounts for the coupling to 
the nuclear system and any external field in an exact way.
 We present two different ways of decomposing this potential and some approximate potentials based on conventional approximations. 
In Sec.~\ref{sec:crei_isotopologues} we study a one-dimensional model for the 
(a)symmetric isotopologues of H$_2^+$ subject to a linearly polarized laser field for two different situations, comparing with dynamics on different approximate potentials.
In Sec.~\ref{sec:quasi-cl} we take a first step to analyse quasiclassically the structure of various components of the potential driving the ionizing electron. 
Finally, some concluding remarks are presented in Sec.~\ref{sec:Conclusions}.

\section{Exact Factorization Approach}
\label{sec:exact_fact}
The exact factorization of the time-dependent electron-nuclear wavefunction introduced in Refs.~\cite{AMG10,AMG12,AMG13,AASG13,SAMYG14} enables the rigorous 
definitions of exact potentials acting on the nuclear subsystem and electronic subsystem 
in coupled electron-ion dynamics. These potentials follow from writing the full molecular wavefunction as a single product 
$\Psi(\dulr,\dulR,t) = \Phi_\dulR(\dulr, t)\chi(\dulR,t)$, or $\Psi(\dulr,\dulR,t) =  \chi_\dulr(\dulR,t)\Phi(\dulr,t)$, where partial normalization conditions on the conditional 
wavefunctions $\Phi_\dulR(\dulr,t)$ and $\chi_\dulr(\dulR,t)$ respectively, render each factorization unique up to a gauge transformation. In the first product the equation for
the nuclear wavefunction, $\chi(\dulR,t)$, follows a TDSE, while in the second product the equation for the electronic factor $\Phi(\dulr,t$) is a TDSE. 
When the electronic dynamics is particularly of interest, as in our present study of ionization, we focus on the the second factorization, and investigate the potential
that appears in the TDSE for $\Phi(\dulr,t)$. 

If we consider the case of one electron coupled to one nuclear degree of freedom in one dimension, the equation for the electronic wavefunction is 
(in  1D we replace $\dulr$ and $\dulR$ by $z$ and $R$ respectively.) :
\ben
\label{eq:eleq-ef}
\left[  \frac{\left(-i\hbar \partial/\partial z+\boldsymbol{\mathcal{S}}(z,t)\right)^2}{2\,m_e}+\epsilon_e(z,t)\right]\Phi(z,t)=i \hbar\,\partial_t\Phi(z,t).
\een
and that for the nuclear conditional wavefunction is:
\ben
\label{eq:exact_ne}
\left[ \hat{H}_{n}(z,R,t)-\epsilon_e(z,t)\right] \chi_z(R,t) = i \hbar \, \partial_t  \chi_z(R,t),
\een
with the nuclear Hamiltonian
\begin{eqnarray}
\hat{H}_{n}(z,R,t) &=& \hat{T}_n (R) + \hat{W}_{en}(z,R) + \hat{W}_{nn}(R) + \hat{v}_{\rm ext}^{n}(R,t) \nonumber \\
&+&  \frac{1}{m_e}  \left[ \frac{[- i \hbar \partial/\partial_z - \boldsymbol{\mathcal{S}}(z,t)]^2}{2} \right.  \nonumber \\
&+& \left. (\frac{-i \hbar \partial\Phi/\partial_z}{\Phi}+ \boldsymbol{\mathcal{S}}(z,t))(- i \partial/\partial_z - \boldsymbol{\mathcal{S}}(z,t)) \right]\nonumber\\
\end{eqnarray}
where $\hat{T}_n$ is the nuclear kinetic energy operator, $\hat{W}_{en}(z,R)$  ($\hat{W}_{nn}(R)$) is  the  electron-nuclear (nuclear-nuclear) interaction, and $\hat{v}_{\rm ext}^{n}(R,t)$ 
is time-dependent external potentials acting on the nuclei.
Here, $\boldsymbol{\mathcal{S}}(z,t)$ is the exact time-dependent vector potential
\begin{equation}
 \boldsymbol{\mathcal{S}}(z,t)= \left\langle\chi_z(R,t) \right\vert -i \hbar \partial/\partial_z \left. \chi_z(R,t)\right\rangle_R 
\end{equation}
and $\epsilon_e(z,t) = \langle\chi_z(t)\vert \hat{H_n} - i\hbar\partial_t\vert\chi_z(t)\rangle_R$ is the exact time-dependent electronic potential for electron (see next section). 
In one dimensional models, $\boldsymbol{\mathcal{S}}(z,t)$ can be
set to zero as a choice of gauge and we adopt this gauge for the rest of this paper. In this case, $\epsilon_e(z,t)$ is  the sole potential that drives the electronic motion, which  can be compared with the
other traditional potentials that are used to study electronic dynamics.

\subsection{Decomposition of the exact time-dependent potential energy surface}
\label{sec:decomp}
The exact time-dependent potential energy surface for electron ($e$-TDPES) contains the effects of the 
coupling to  moving quantum nuclei as well as the external laser field. It can be written as
\ben
\label{eq:exact_etdpes}
\epsilon_e(z,t) = \epsilon^{\rm app}_e(z,t) + \mathcal{T}_n (z,t)+ \mathcal{K}_e^{\rm cond}(z,t) +  \epsilon^{\rm gd}_e(z,t),
\een 
which consists of: the approximate potential
\ben
\label{eq:eps_approx}
\begin{split}
&\epsilon^{\rm app}_e(z,t)= \left\langle \chi_z(R,t)|\hat{v}\ext^{e}(z,t)|  \chi_z(R,t)\right\rangle_R \\
 &+\left\langle\chi_z(R,t) \right\vert\hat{W}_{en}(z,R)+\hat{W}_{nn}(R) \left\vert \chi_z(R,t)\right\rangle_R, 
 \end{split}
\een
the nuclear-kinetic term 
\ben
\mathcal{T}_n(z,t) = -  \frac{ \hbar^2}{2 m_n}\left\langle\chi_{z}(R,t) \right\vert \frac{\partial^2}{\partial R^2}  \left\vert \chi_{z}(R,t)\right\rangle_R,
\label{eq:kinetic}
\een
 the gauge dependent part of the potential
\ben
\epsilon^{\rm gd}_e(z,t) = \hbar \left\langle\chi_{z}(R,t)\right\vert - i \partial_t\left. \chi_{z}(R,t)\right\rangle_R\nonumber,
\een
and finally the electronic-kinetic-like contribution

$$\mathcal{K}_e^{\rm cond}(z,t)=  \frac{\hbar^2}{2 m_e}\ {\left\langle \frac{\partial}{\partial z}  \chi_{z}(R,t) \vert \frac{\partial}{\partial z}  \chi_{z}(R,t)
  \right\rangle_R } .$$
In Ref.~\cite{KAM15}, we had found that this exact potential $\epsilon_e(z,t)$,  has significant features that are missing in the traditional 
potentials based upon the quasistatic picture described above. Neglecting these features led to qualitatively
incorrect predictions of ionization dynamics in the H$_2^+$ molecule
in strong fields. Ref.~\cite{KAM15} found that, in general, all the four terms above are needed to reasonably reproduce the ionization in CREI processes. i.e. that propagation on any combination other than the full sum of the four contributions in Eq.~\ref{eq:exact_etdpes} gave qualitatively poor results. This means that in eventually developing approximations, all of the four terms must be considered.

Alternatively, one may instead decompose the $e$-TDPES by exactly inverting the TDSE for the electronic wavefunction.
The electronic wavefunction may be written in polar representation, 
\ben
 \Phi(z,t) = \sqrt{n_e(z,t)} \exp{(i \alpha(z,t))}
\een
 with $n_e(z,t)  = \vert\Phi(z,t)\vert^2$ and $\alpha(z,t) = \frac{m_e}{\hbar} \int^{z} dz' \frac{{\bf j_e}(z',t)}{n_e(z',t)}$ where 
 \ben
 {\bf j_e}(z,t) =  \frac {\hbar}{m_e} \Im ( \Phi^{*}(z,t) \nabla_{e} \Phi(z,t)),
 \een
 is the electronic current-density.
Inserting this form into Eq.~\ref{eq:eleq-ef} and setting the time-dependent vector potential to zero, gives
\ben
\label{eq:exact_etdpes_tddft}
\epsilon_{e}(z,t)=  \frac{\hbar^2}{2 m_e} \left[ \frac{{\nabla^{2}_{e}}\sqrt{n_e(z,t)}}{\sqrt{n_e(z,t)}}\right]  -  \frac{m_e}{2}\left(\frac {{\bf j_e} (z,t)}{n_e(z,t)}\right)^2- \partial_t  \alpha(z,t)
\een

The first term is what survives in the absence of any dynamics, and it
depends only on the instantaneous electron density; we denote it
``adiabatic'' in the spirit of time-dependent density functional
theory.  The second term, denoted ``velocity term'', depends only on
the electron velocity, namely on ${\bf j_e}(z,t)/n_e(z,t)$, while the last
term, denoted ``acceleration term'' depends on the spatial integral of
the acceleration. In Section~\ref{sec:cpsir}
 and~\ref{sec:marginal-structure}
, we consider propagation on
different contributions of the exact potential defined by the
decomposition of Eq.~\ref{eq:exact_etdpes_tddft}, again with a view to 
eventually developing approximations, possibly density functionals (see Ref.~\cite{RG16}), for the exact $e$-TDPES. 

\subsection{Approximate electronic potentials based on conventional approximations}
\label{sec:app-pot}
The standard approximation for the electronic potential assumes the so-called quasistatic approximation (qs) that treats the nuclei as classical point particles with positions 
that are either considered fixed , $\bar{R}_0$, as in the majority of studies of CREI~\cite{ZB95,CFB98,CB95,CCZB96,YZB95,BL12}, or move classically
with classical trajectories $\bar{R}(t)$ that are often described by
mixed quantum-classical algorithms such as Ehrenfest or
surface-hopping algorithms \cite{UGKS07}. In these methods, electrons on the other hand,
regardless of whether the nuclei are frozen or move classically, follow
the combined potential from the laser field and 
the electrostatic attraction of the nuclei, i. e.,
\ben
\label{trad_tdpes}
\epsilon^{\rm qs}(z,t|\bar{R}(t))= \hat{W}_{en}(z,\bar{R}(t)) +\hat{V}_e^{l}(z ,t).  
\een 
Considering the exact electronic potential Eq.~\ref{eq:exact_etdpes}, we see that such approaches completely miss the dynamical electron-nuclear 
correlation effects contained in $\mathcal{T}_n(z,t)$, $\mathcal{K}_e^{\rm cond}(z,t)$, and $\epsilon^{\rm gd}$, and can be viewed as an 
{\it approximation} to $\epsilon^{app}$ alone: $\epsilon^{\rm app}$ reduces to the qs approximation when the conditional nuclear wavefunction 
is approximated classically as a $z$-independent delta-function at $\bar{R}(t)$ , i.e. $ n_{z}(R,t)
\approx \delta(R(t),\bar{R}(t))$ (in this limit
$\hat{W}_{nn}(\bar{R}(t))$ becomes purely a time-dependent constant and
hence is dropped hereafter). 

A step beyond the qs approximation for the electronic potential that can, in principle, account for the width 
and splitting of the nuclear wavefunction is the electrostatic or Hartree approximation \cite{T98}
\ben
\label{trad-H}
\epsilon_e^{\rm Hartree}(z,t) = \hat{V}_e^{l}(z,t) + \int \, dR  \hat{W}_{en}(z,R) n(R,t),
\een
where $n(R,t)$ is the nuclear density obtained from nuclear wavepacket dynamics
\footnote{The Hartree potential~(\ref{trad-H}) reduces to the qs
  expression~(\ref{trad_tdpes}) in the limit of very localized
  wave-packets centered around $\bar{R}$.}. It can be
easily seen that if the  $z$-dependence in the conditional nuclear
wavefunction is neglected, i. e., $ n_{z}(R,t) \approx n(R,t)$, the
approximate potential simplifies to the Hartree approximation.

To provide a detailed analysis of the exact electronic potential, we
compare the electronic dynamics resulting from different approximations. We will consider
combinations of the terms of the exact potential decomposed
according to Eq.~\ref{eq:exact_etdpes_tddft} to complement the analysis in Ref.~\cite{KAM15} given in terms of the decomposition in Eq.~\ref{eq:exact_etdpes}, as well as
the electronic dynamics resulting from the following approximations: i) The qs approximation for which
$\bar{R}(t)=\langle R \rangle (t)$ is the average time-dependent internuclear
separation obtained from the exact 
calculations. ii) The Hartree approximation for which the nuclear
density in (\ref{trad-H}) is replaced by the exact  time-dependent nuclear density. Hence
we write the corresponding electronic potential as $\epsilon^{\rm ex-H}$
to indicate that the exact nuclear density is substituted into the
Hartree expression. iii) Due to a considerable loss of norm in the
CREI regime, we normalize the electrostatic part of the potential in 
Eq.~\ref{trad-H} to obtain normalized Hartree as
\ben
\label{trad-H-norm}
\epsilon^{\rm n-Hartree}(z,t)=\hat{V}_e^{l}(z,t) + \frac{\int \, dR  \hat{W}_{en}(z,R) n(R,t)}{\int \, dR \, n(R)},\
\een
indicated  as  $\epsilon^{\rm n-ex-H}$ when the exact nuclear
  density is inserted in Eq.~\ref{trad-H-norm}.
 iv) the ``self-consistent'' Hartree approximation (SCH) in which the
full wavefunction is approximated as an uncorrelated product of electronic
wavefunction and nuclear wavefunction, $\Psi_H(\dulr,\dulR,t) =
\phi(\dulr,t)\chi(\dulR,t)$ that treats the electrons and nuclei on the
same footing:  the electronic dynamics is described with the potential
in Eq.~\ref{trad-H} that is coupled to the nuclei that move under
the influence of the analogous potential, i.e.,    
\ben
\label{trad-H-n}
\epsilon_n^{\rm Hartree}(R,t) = \hat{V}_n^{l}(R,t) + \int \, dz  \hat{W}_{en}(z,R) \n_e(z,t),
\een
where $\n_e(z,t)$ is the electron density obtained from the electronic
dynamics. This approach does not involve the complications of dealing with the conditional nuclear wavefunction as in $\epsilon^{\rm app}$
but it fails to capture major effects of the correlated
electron-nuclear dynamics due to its mean-field nature. Finally, we point out that $\epsilon^{app}$ as well as all the approximations (i)--(iii) 
represent the electron-nuclear interaction in an electrostatic way only.
 
\section{CREI in H$_2^+$ Isotopologues}
\label{sec:crei_isotopologues}
We utilize a popular one-dimensional model of the symmetric as well as asymmetric isotopologues of   H$_2^+$ subject to a linearly polarized laser field.
As the motion of the nuclei and the electron in the true molecule is assumed to be restricted to the direction of the polarization axis of the laser field,
 the essential physics can be captured by a 1D Hamiltonian featuring ``soft-Coulomb'' interactions~\cite{JES88, KLEG01}:
\begin{eqnarray}
  \label{eq:Hamiltonian}
    \hat{H}(t) &=&  - \frac{\hbar^2}{2\mu_e}\frac{\partial^2}{\partial z^2} - \frac{\hbar^2}{2\mu_n}\frac{\partial^2}{\partial R^2}- \frac{1}{\sqrt{1+(z-\frac{M_2}{M_n} R)^2}} \nonumber \\
       &-&  \frac{1}{\sqrt{1+(z+\frac{M_1}{M_n} R)^2}} + \frac{1}{\sqrt{0.03+R^2}} +  \hat{V}_l(R,z,t) \nonumber \\
       &&
\end{eqnarray}
\begin{figure}
\begin{center}
\includegraphics[width=0.5\textwidth]{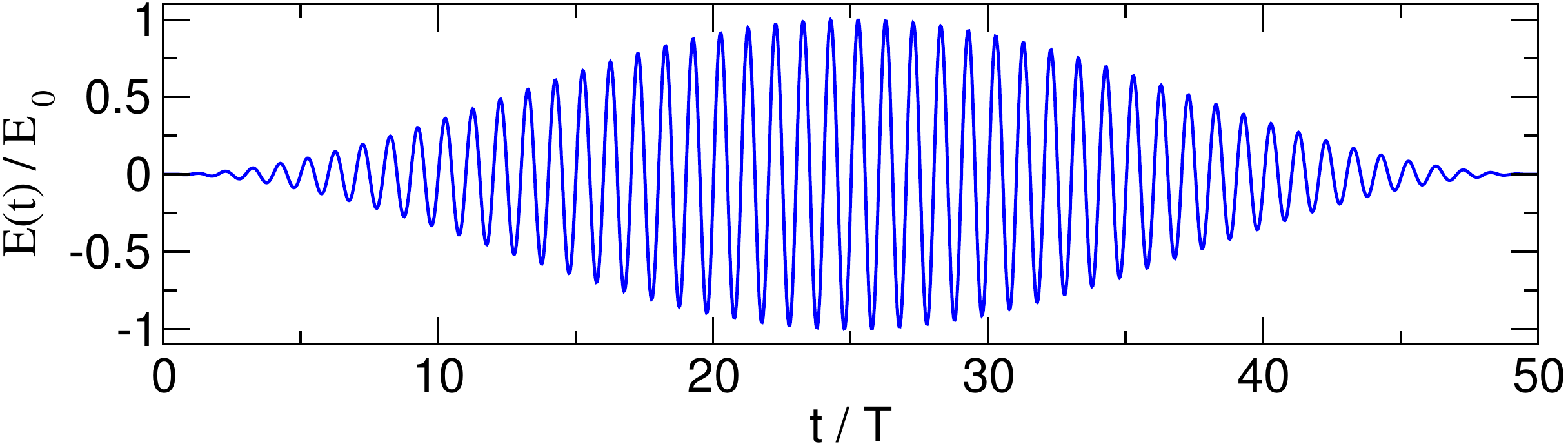}
\end{center}
\caption{Laser field as a function of number of optical cycles $t/T$. The
  electric field amplitude is divided by the peak amplitude, $E_0 =\sqrt{I}$. }
\label{fig:laser}
\end{figure}
where $R$ and $z$ are the internuclear distance and the electronic
coordinate as measured from the nuclear center-of-mass, respectively.
The nuclear effective mass is denoted as $\mu_n = \frac{M_1 M_2}{M_n}$ while $\mu_e=\frac{M_n}{M_n + m_e}$  is the electronic reduced mass with $M_n\,=\,M_1 + M_2$. 
The laser field, within the dipole approximation, is represented by
\begin{equation}
\label{eq:dipole}
\hat{V}_l(R,z,t) =  e E(t) (q_e\, z - \zeta R),            
\end{equation}
 where $E(t)$ denotes the electric field
amplitude and  $q_e =1 	  + \frac{m_e}{M_n+m_e}$ is the reduced charge and $\zeta = (M_2-M_1)/M_n $ is the mass-asymmetry parameter. 
Such reduced-dimensional models  have been shown to qualitatively
reproduce experimental results (see Ref.~\cite{KMS96} for example).

We study the symmetric isotopologues, i.e.,  H$_2^+$, A$_2^+$, D$_2^+$, x$_2^+$ 
as well as the asymmetric isotopologues HD$^+$, HT$^+$, Hx$^+$, HX$^+$.   Here x(X)  stands for the fictitious isotope of hydrogen that 
is 10(100) times heavier than that of H, while A$_2^+$ is another fictitious isotopologue with the same effective nuclear mass as HT$^+$ (See Table.~\ref{tab:eff-mass} in which the nuclear effective mass of H$_2^+$ and  isotopologues are given.)
\begin{table}

 \begin{tabular}{| c | c | c | c | c | c | c | r |}
  \hline
  
   H$_2^+$ & HD$^+$ & HT$^+$(A$_2^+$) & Hx$^+$ & HX$^+$ & D$_2^+$ & x$_2^+$ \\
  \hline 
   918.076 & 1223.742 & 1376.228 & 1669.229 & 1817.973 & 1834.533 & 9180.76\\
  \hline
  \end{tabular}
 
\caption{Nuclear effective mass corresponding to H$_2^+$ and its symmetric and antisymmetric isotopologues in atomic units. Here, x(X) refers
to the 10(100) times heavier fictitious isotope of hydrogen and A$_2^+$ is another fictitious isotopologues with the same effective nuclear mass as HT$^+$.  \label{tab:eff-mass}}
\end{table}

\subsection{Comparison of ionization of isotopologues of H$_2^+$ with different effective nuclear mass for a fixed field}
\label{sec:sfdi}
We apply a field of duration 50-cycles, wavelength $\lambda\,= 800$ nm ($\omega=0.0569$ a.u.) 
and intensity $I = 2.02 \times 10^{14}$W/cm$^2$, with a sine-squared
pulse envelope (Fig.~\ref{fig:laser}), to each of the H$_2^+$ isotopologues.
First, we solve the electron-nuclear TDSE for Hamiltonian of Eq.~\ref{eq:Hamiltonian}
numerically exactly, beginning in the initial ground-state of the molecule. 

\begin{figure}
\begin{center}
\includegraphics[width=0.5\textwidth]{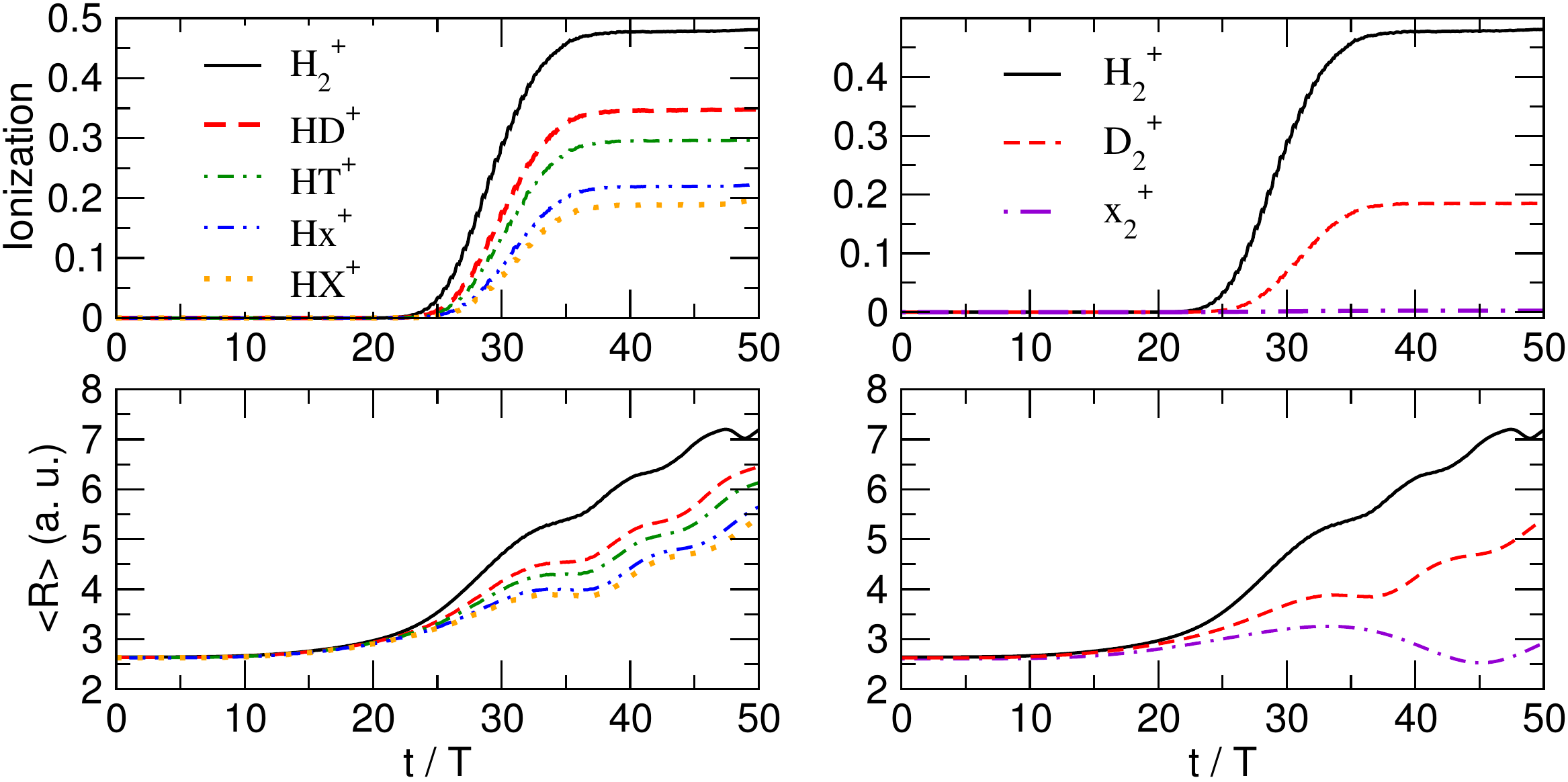}
\end{center}
\caption{Ionization probability and average internuclear distance ,$\langle R \rangle $, as a function of number of cycles $t/T$ where $T$ denotes 
duration of one cycle ($T=2.67$ fs)  of the H$_2^+$, HD$^+$, HT$^+$, Hx$^+$ and HX$^+$ molecule on the left panels and A$ _2^+$, D$ _2^+$ and x$ _2^+$ on the right panels.}
\label{fig:samefield_diffM}
\end{figure}


As the Hamiltonian~(\ref{eq:Hamiltonian}) has been obtained after separating off the center of mass motion and the origin is set to be the nuclear center of mass the field couples directly to the nuclear motion only in the asymmetric cases; in the symmetric case, nuclear motion is driven 
purely by the electronic dynamics. This is also clear from Eq.~\ref{eq:dipole} where in the symmetric case, $\zeta = 0$.
In Figs.~\ref{fig:samefield_diffM}  we plot the ionization probability and average internuclear distance ,$\langle R \rangle $, as a function of 
number of cycles $t/T$ where $T$ denotes duration of one cycle ($T=2.67$ fs) for H$_2^+$, HD$^+$, HT$^+$, Hx$^+$,  HX$^+$ , A$ _2^+$, D$ _2^+$ 
and x$_2^+$. Note that, the results are given in atomic units, $e = m_e = h= 1$, throughout
the article, unless otherwise noted. The ionization probability is defined as $I(t) =1 - \int_{-\infty}^{\infty} dR \int_{-z_I}^{z_I} dz \vert\Psi(z,R,t)\vert^2$,
with  $z_{I} = 15$ a.u. 
\begin{figure}
\begin{center}
\includegraphics[width=0.5\textwidth]{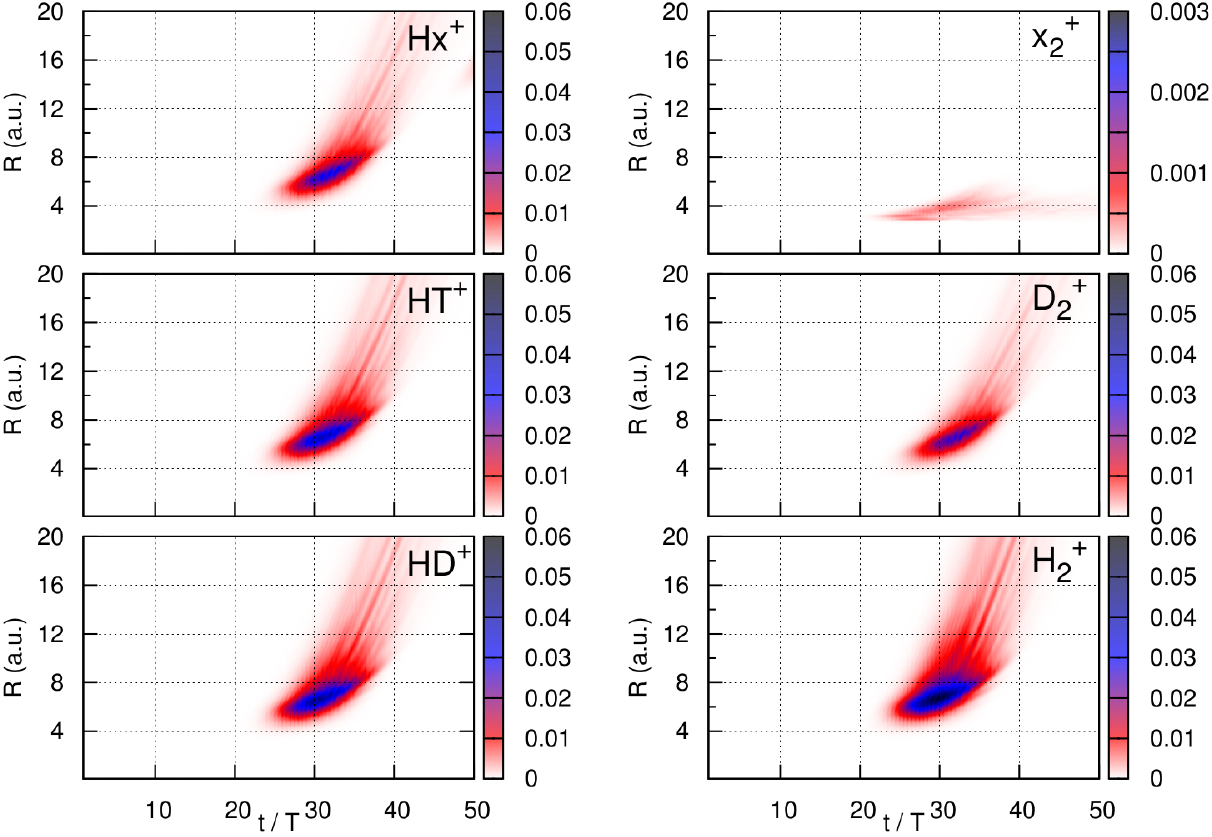}
\includegraphics[width=0.5\textwidth]{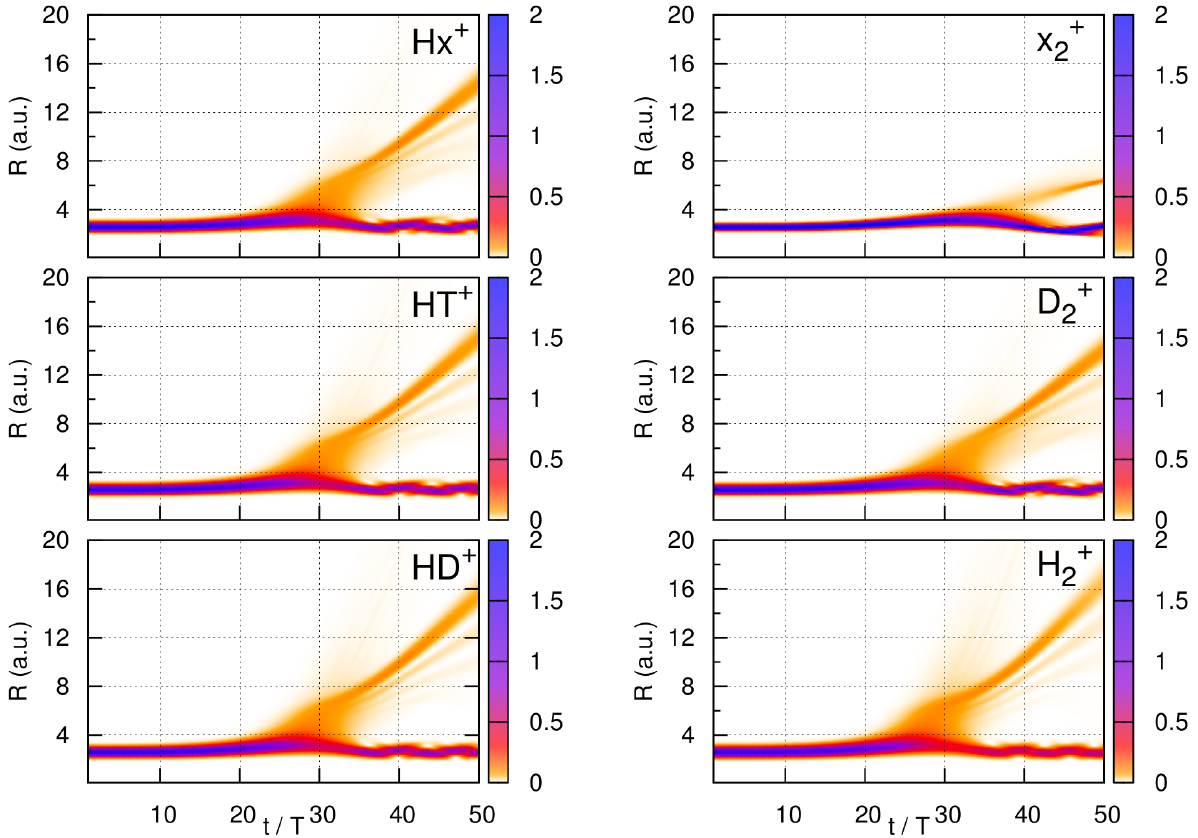}
\end{center}
\caption{The time-resolved, $R-$resolved ionization probability,
  $I(R,t)$, (upper panels) and the nuclear density (lower panels) as a function of number of cycle for asymmetric case in the left panels
HD$^+$, HT$^+$, Hx$^+$ and symmetric case in the right panels H$ _2^+$, D$ _2^+$ and x$ _2^+$.}
\label{fig:samefield_diffM_nIR}
\end{figure}
As it can be seen in Figs.~\ref{fig:samefield_diffM}, the ionization and average internuclear separation appear to be
practically independent of nuclear mass-symmetry properties: they are identical for asymmetric case of HT$^+$ and symmetric
case of A$ _2^+$ systems with the same effective nuclear mass $\mu_n$ and very similar in the case of HX$^+$ and  D$ _2^+$ where the effective masses are only a little different.
One can understand this from  that fact that the ionization probes the electron density in regions where $z \gg R$, where the nuclear mass-dependence in the Hamiltonian 
Eq.~\ref{eq:Hamiltonian} to lowest order in $R/z$ is only via $\mu_n$, and the integrated nature of the observable reduces its sensitivity to the details of the distribution 
which is affected by the symmetry of the system. 

It is also observed that as the nuclear effective-mass increases, the
ionization decreases, with a dependence that appears to tend to $1/\mu_n$ for
intermediate masses. Furthermore, the growth over time of  the average internuclear distance is less
as the mass increases, suggesting that for larger masses the system reaches the critical internuclear distance for CREI at later times when the field 
has decreased from its peak value significantly, consequently resulting in lower ionization. 
To investigate whether the CREI process is still
  relevant or not and shed more light on the dependence of the 
ionization on the effective nuclear mass we utilize the concept of the time-resolved, $R-$resolved ionization probability defined as $I(R,t) =\int_{z'_I} dz \vert\Psi(z,R,t)\vert^2$,
with  $\int_{z'_I} = \int_{-\infty}^{-z_{I}} +\int^{\infty}_{z_{I}}$ and $z_{I} = 15$ a.u. ~\cite{KAM15,KMS96}. This quantity can be rewritten in terms of the 
concepts of exact nuclear wavefunction $\chi(R,t)$ and exact conditional electronic wavefunction $\Phi_R(z, t)$ introduced within the exact factorization framework, i. e., 
\ben
\label{eq:dwcip}
I(R,t)= |\chi(R,t)|^2 I_{cp}, 
\een 
where, $I_{cp}$ follows the usual expression of the ionization probability
but using the exact conditional electronic wavefunction $\Phi_R(z,t)$ , i.e., $I_{cp} (R,t)=1-\int_{-z_{I}}^{z_{I}} dz \vert\Phi_R(z, t)\vert^2$
that is coupled to the exact nuclear dynamics. {\it Therefore,} $I(R,t)$, {\it which is the nuclear density weighted conditional ionization probability is analogous to the ionization probability calculated 
for a given nuclear configuration $R$ in quasi-static picture.}

To analyze the dynamics of different cases, in
Fig.~\ref{fig:samefield_diffM_nIR} we have plotted $I(R,t)$ along with the time-dependent nuclear density for different isotopologues. As seen in the upper three rows of Fig.~\ref{fig:samefield_diffM_nIR} with increasing the effective nuclear mass the peak of $I(R,t)$ 
moves slightly to larger times and its overall value decreases while remaining close to the CREI region as predicted by the internuclear separation of $R_c$. The lower set of panels of 
Fig.~\ref{fig:samefield_diffM_nIR} shows the time-dependent nuclear density in each case. For the case of x$_2^+$ where only an exponentially small fraction of the nuclear density 
dissociates, the system does not reach the CREI regime and therefore 
the ionization is negligible as seen from the $R-$resolved, $t-$resolved ionization probability. Notice the different scale 
of $I(R,t)$ for the case of x$_2^+$. 

Note that the ionization rate computed at any fixed $R$ would be
{\it identical} for all the isotopologues. Their different masses, however, lead to
very different ionization rates when the full electron-nuclear
dynamics is considered with the molecule beginning  at equilibrium as
we have shown. Still, there is some validation to the original
quasistatic CREI prediction that ionization is enhanced for nuclear
separations around $R_c$, as indicated by $I(R,t)$, however, a
modification to the statement is needed due to the spreading and
splitting of the nuclear wavepacket: the enhancement occurs from
electrons associated with the part of the nuclear density that is in
the $R_c$ region. Clearly, treating the nuclei as point particles
will not work even for significant nuclear mass (except in the
large-mass limit) because of this. The question then arises whether accounting for the nuclear distribution is enough to capture CREI accurately, for example using one of the electrostatically-based approximations of Section~\ref{sec:exact_fact}
, and, whether and how the errors decrease with the nuclear mass. To this end, we next consider adjusting the field strength so that the ionization is similar for the different isotopologues.

\subsection{Comparison of potentials with similar ionization rates}
\label{sec:cpsir}

\begin{table}
 \begin{tabular}{| c | c | c | c |}
  \hline
   H$_2^+$ & A$_2^+$ & D$_2^+$ & x$_2^+$\\
  \hline 
  $ 2.02 \times 10^{14}$ &  $2.14 \times 10^{14}$  &  $2.26 \times 10^{14}$ &  $3.1 \times10^{14}$  \\
  \hline
  \end{tabular}
\caption{The optimized field intensity for different-mass systems in the unit of W/cm$^2$.  \label{tab:opt-int}  }
\end{table}
As discussed in the previous section, the different isotopologues of H$_2^+$ subject to the same field 
show significantly different degrees of ionization, since the different-mass systems reach the 
CREI region at different times with different probabilities. 
In particular, systems with larger nuclear masses hardly reach the internuclear distances for which CREI is expected to occur.
Hence, to study the effect of the nuclear mass in the CREI regime for different-mass systems, we adjust the field intensity such that the ionization probability(rate)
remains close to that of H$_2^+$. As the asymmetric and symmetric isotopologues with the same effective nuclear mass subject to the same field give
 almost the same ionization probability, average internuclear distance and $I(R,t)$ (see section \ref{sec:sfdi}), from here on we focus only on the symmetric isotopologues of H$_2^+$.
 The optimized field intensity for different-mass systems is given in 
Table.~\ref{tab:opt-int} while the other field parameters are kept unchanged. 

\begin{figure}
\begin{center}
\includegraphics[width=0.45\textwidth]{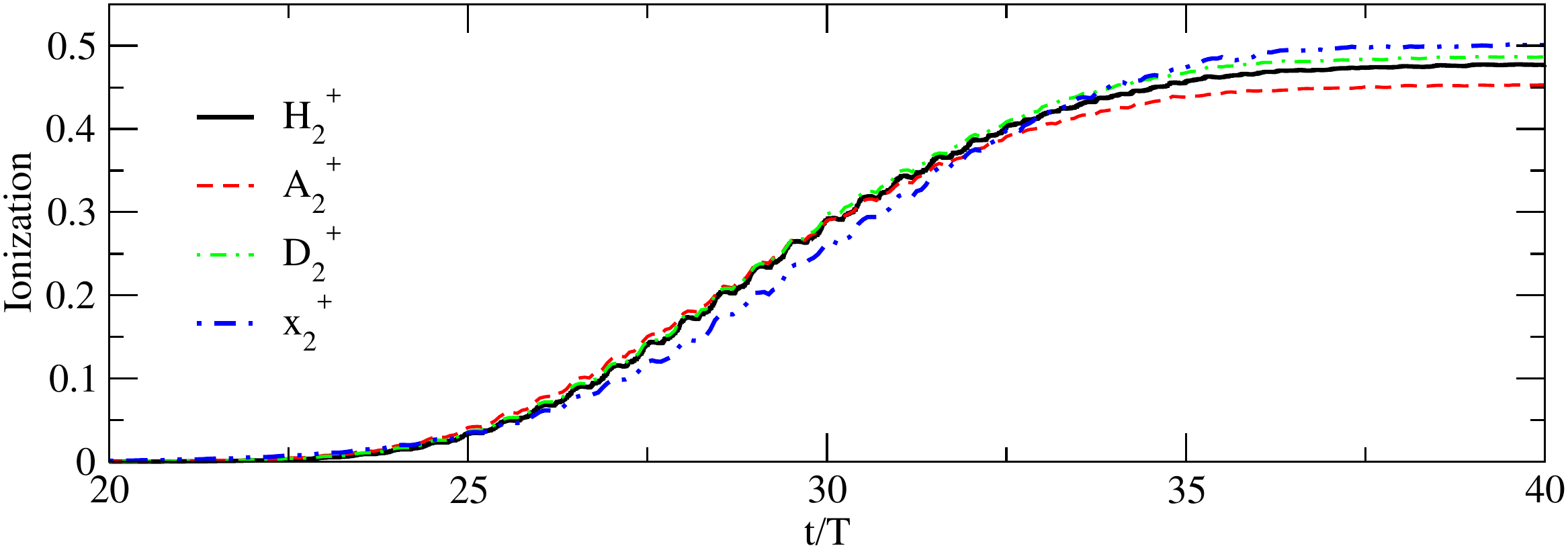}
\end{center}
\caption{The field intensity is varied such that the ionization yield(rate) of different systems remains similar to that of H$_2^+$ subject to 
a $50$-cycle pulse of wavelength $\lambda\,= 800$ nm ($\omega=0.0569$ a.u.) and intensity $I = 2.02 \times 10^{14}$W/cm$^2$, with a sine-squared pulse 
envelope. For the isotopologues of H$_2^+$, to view better how the ionization yield(rate) compares to that of H$_2^+$ we have shifted the ionization 
by one cycle ( i.e. I((t-T)/T) is plotted).  }
\label{fig:opt-ion}
\end{figure}

The ionization probability of the symmetric isotopologues subject to
the optimized field is depicted in Fig.~\ref{fig:opt-ion}. For
isotopologues heavier than H$_2^+$  the ionization starts to set in about one optical cycle $T$ later than the H$_2^+$  
case. In order to be able to compare the ionization yield (rate) of these systems with H$_2^+$ {\it visually} better, in Fig.~\ref{fig:opt-ion} we have shifted the time-dependent ionization probability by one 
optical cycle, i.e. for isotopologues heavier than H$_2^+$ , $I(t-T)$ has been plotted.

\begin{figure}
\begin{center}

\includegraphics[width=0.5\textwidth]{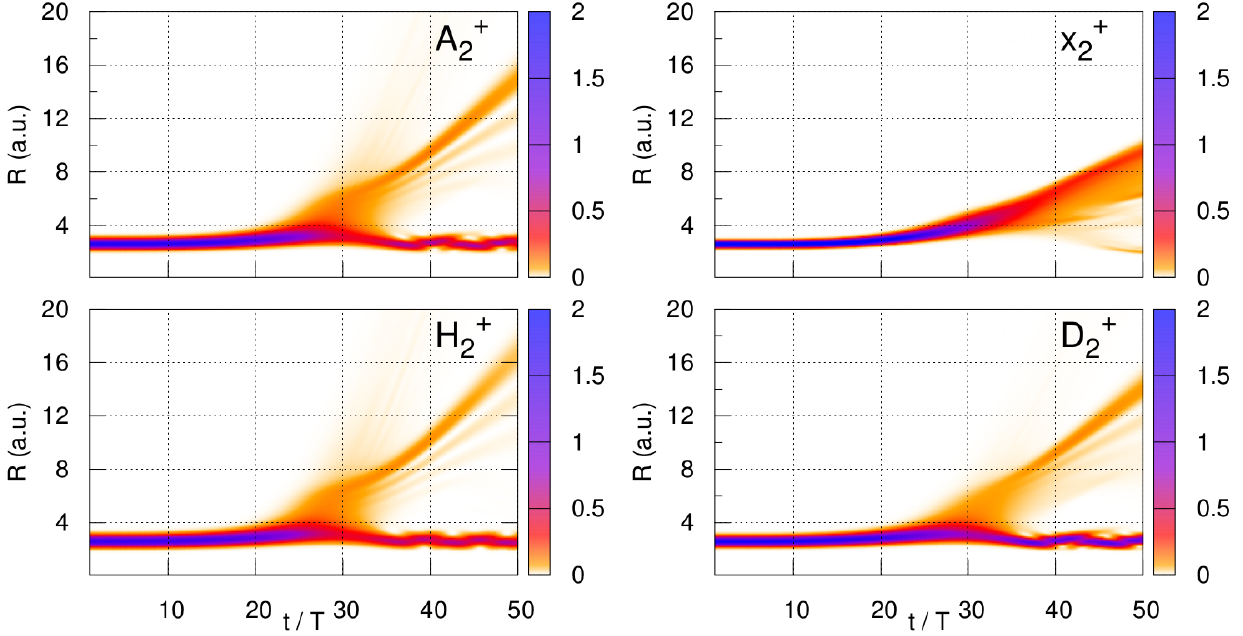}
\end{center}
\caption{The contour plot of the time-dependent nuclear density for H$_2^+$ (lower left panel),  A$_2^+$ (upper left panel)
 , D$_2^+$ (lower right panel),  x$_2^+$ (upper right panel) subject to the optimized field given in table \ref{tab:opt-int}. }
\label{fig:sidf-nd}
\end{figure}

\begin{figure}
\begin{center}

\includegraphics[width=0.5\textwidth]{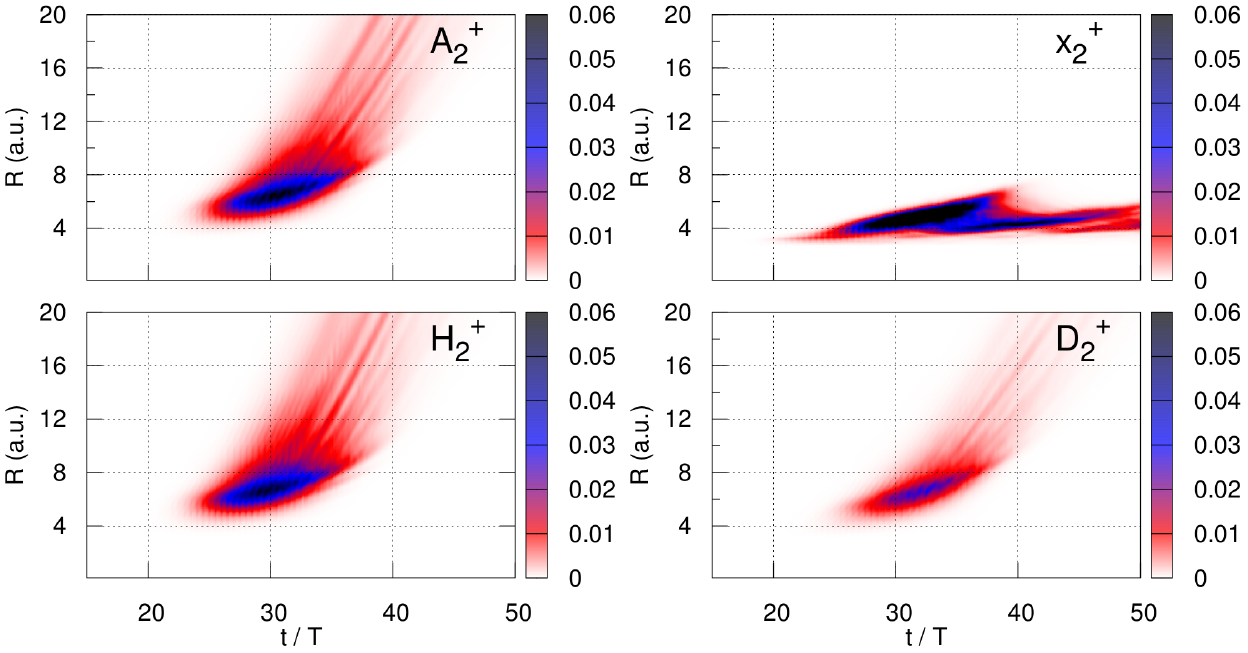}
\end{center}
\caption{The contour plot of $I(R,t)$  for H$_2^+$ (lower left panel),  A$_2^+$ (upper left panel)
 , D$_2^+$ (lower right panel),  x$_2^+$ (upper right panel) subject to the optimized field given in table \ref{tab:opt-int}. }
\label{fig:sidf-IR}
\end{figure}

To analyze the mechanism of ionization in more detail, in Fig.~\ref{fig:sidf-nd} we plot the time-dependent nuclear density for the
various isotopologues. The nuclear density behaves similarly in all
cases, except the heaviest case, namely x$_2^+$. That is 
before the field reaches its maximum the system becomes slightly
  ionized hence the nuclear density slightly spreads. As the field
  reaches its maximum a small fragment of  
the nuclear density starts to split off and dissociate from Coulomb explosion due to an increase
in the ionization while the rest of the nuclear density remains
bounded and oscillates around the equilibrium position. 
As an appreciable amount of dissociating fragment reaches the
internuclear separation for CREI, the ionization gets enhanced. In the case of x$_2^+$, however, the nuclear dynamics exhibits a  
more classical behavior, i. e. during the first half of the pulse, the heavy nuclei only
spreads slightly around the equilibrium. The stronger field
compared to the other cases enables the system to ionize initially (before the nuclear density reaches the critical $R$). This initial ionization will be
followed then by a Coulomb explosion toward the middle of the field,
after which the nuclear density spreads further but it hardly
splits. The remaining part of the electronic density undergoes
enhanced ionization as the nuclear density lies in the range of the
internuclear separation associated to CREI.
In this case, the average internuclear distance almost coincides with the peak of the nuclear wave packet and therefore for 
systems with very large effective nuclear mass the standard approximation to describe CREI, namely the qs approximation is expected to perform better 
compared to the lighter isotopes. 
The different nature of ionization in large nuclear-mass systems, can also be seen from  $I(R,t)$ depicted in Fig.~\ref{fig:sidf-IR}. 
While for not too heavy isotopologues namely A$_2^+$ and D$_2^+$ , the $I(R,t)$ has a similar structure to H$_2^+$, the $I(R,t)$ 
corresponding to x$_2^{+}$ manifests a substantially distinctive structure compared to the lighter isotopologues:
the internuclear separation for which ionization gets enhanced, shifts
to smaller values. In general, the $I(R,t)$ shifts to smaller $R$ for heavier isotopologues which could be related to 
the stronger optimized field used \cite{CB95,CFB98}.    


\begin{figure}
\begin{center}
\includegraphics[width=0.5\textwidth]{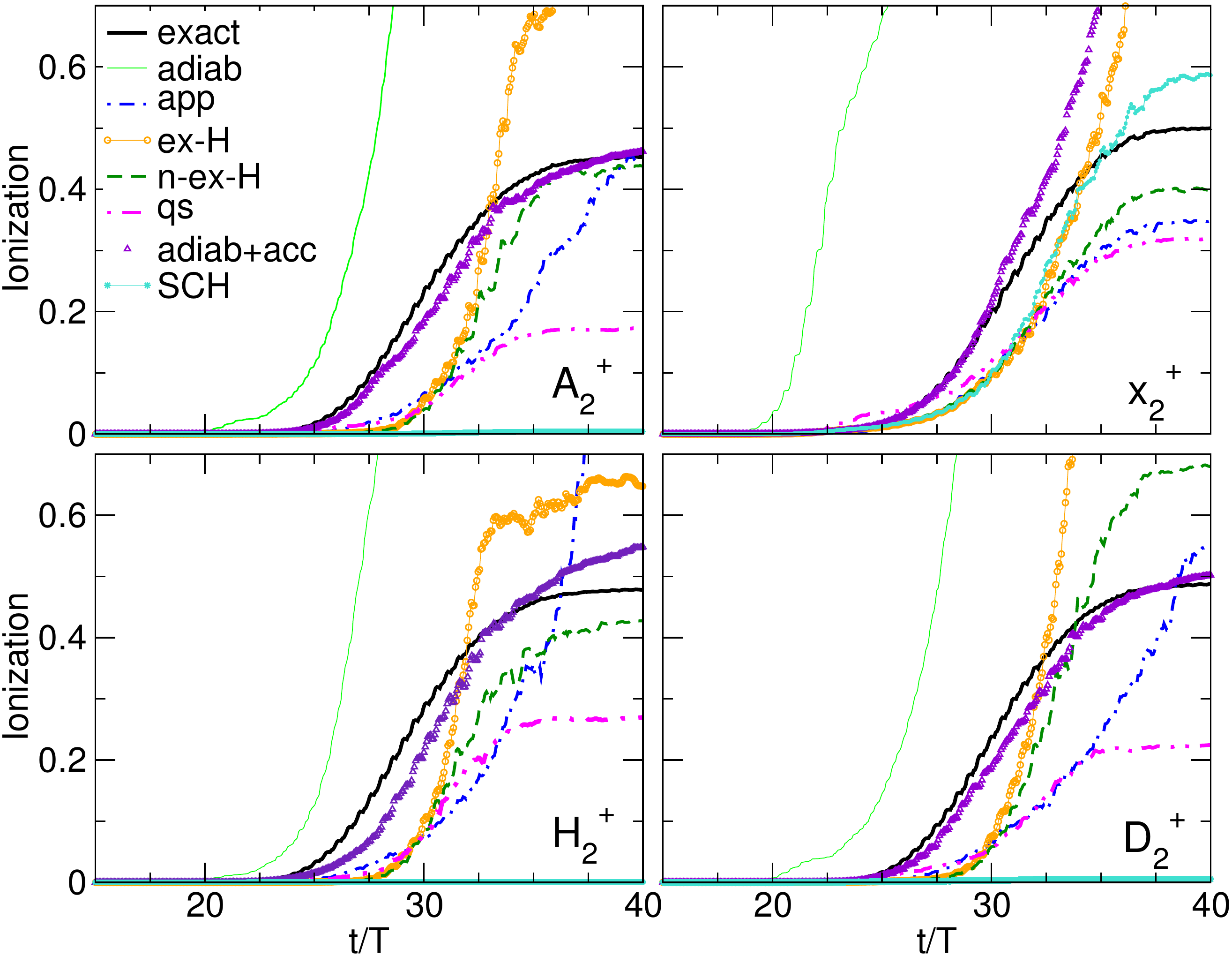}
\end{center}
\caption{ Ionization probabilities calculated from propagating the electron on  the exact, adiabatic, approximate, Hartree like, normalized Hartree and quasi-static
PES for different isomer of ionic Hydrogen.  }
\label{fig:sidf-comp-app}
\end{figure}

Now, we investigate the performance of the various approximations introduced in Sec.~\ref{sec:app-pot} for different-mass systems. In Fig.~\ref{fig:sidf-comp-app} 
the ionization probabilities calculated from propagating the electron on the exact, adiabatic(adiab), approximate(app), exact-Hartree(ex-H), normalized Hartree(n-ex-H), quasi-static(qs) and self-consistent Hartree 
(SCH)
for various symmetric isotopologues of H$_2^+$ are plotted. 

For all cases presented in Fig.~\ref {fig:sidf-comp-app}, propagation
on the qs potential gives rise to an underestimation of the ionization
probability. The average $\langle R \rangle$ entering into the qs potential is always considerably less than the internuclear separation of the dissociating fragment, and so doesn't access the CREI region that long during the duration of the pulse.
The exact-Hartree (which compared to the qs approximation accounts for the spreading and splitting of the nuclear wave packet) follows the qs results until the middle of propagation time 
then overtakes the qs ionization and finally gives a rather large overestimation of the final ionization for all cases. This overestimation can be improved significantly by 
using the normalized Hartree approximation as defined in Eq.~\ref{trad-H-norm}.
In fact, the n-ex-H approximation performs clearly better than other conventional approximations (qs, SCH, ex-H) for H$_2^+$, A$_2^+$ and x$_2^+$. 
Even the approximate potential that depends on the more rigorous and complicated concept of conditional nuclear density rather than the nuclear density that appears in the Hartree-like approximation, 
is considerably worse than n-ex-H. 
The SCH performs very poorly with a negligible ionization probability for not too heavy isomers, 
and only in the large mass limit shows some improvement in predicting the ionization probability, yielding a similar performance to the other conventional approximations.
This very poor performance of the SCH stems from the fact that the SCH is an uncorrelated approach and the splitting of nuclear wave-packet in the CREI regime cannot be accounted for. Hence, for the cases where the CREI occurs due to the 
splitting of a dissociating fragment of nuclear wave packet the SCH cannot capture the right physics. In the large mass limit, however, 
the CREI mechanism relies on the spreading of the nuclear wave packet rather than splitting (see Fig.~\ref{fig:sidf-nd}) , therefore
the SCH approximation  is not as poor. In general, the approximations
do {\it not} perform better with increasing nuclear mass: dynamical
correlation effects that are missing in all the potentials shown
(apart from adiab and adiab+acc) depend more critically on the nuclear
{\it velocities} relative to the electronic, and adjusting the field
to get similar ionization results in these being similar (See also
Sec~\ref{sec:qs-anal-eTDPES}). 

The two curves in Fig.~\ref{fig:sidf-comp-app}  remaining to be discussed are adiab and adiab+acc, which are components arising from the marginal decomposition (Eq.~\ref{eq:exact_etdpes_tddft}). 
The propagation of the electron on the adiabatic potential alone (adiab), gives rise to the complete ionization of the system rather early 
for all isotopologues, while the propagation on the potential composed of the adiabatic plus acceleration (adiab + acc) term yields an ionization probability 
in a very good agreement with the exact results with the velocity term adding only a small correction, expect in the large mass limit. In the following section by studying the structures of the terms 
of the marginal decomposition we try to shed some light on the reason behind the performance of these potentials.


\subsection{The structure of the dynamical electron-nuclear terms in $e$-TDPES: marginal decomposition}
\label{sec:marginal-structure}


We concluded the previous section by briefly discussing the electron dynamics on different terms/combinations-of-terms of the marginal 
decomposition of $e$-TDPES (Eq.~\ref{eq:exact_etdpes_tddft}). In order to better understand the outcomes, in this section we study structures of 
different terms of marginal decompositions of the $e$-TDPES for the two radically different cases of H$_2^+$ and x$_2^+$. 
We refer the readers to Ref.~\cite{KAM15} for a discussion on the components of the conditional decomposition of $e$-TDPES. For the sake of simplifying 
the discussion, here, we divide the electronic coordinate into two regions: "inner-region" that refers to the region with $|z|< 5$ a.u. and 
"outer-region" that describes the rest of the axis for which $|z|> 5$ a.u.
\subsubsection{\it H$_2^+$ case}
\begin{figure}
\begin{center}
\includegraphics[width=0.5\textwidth]{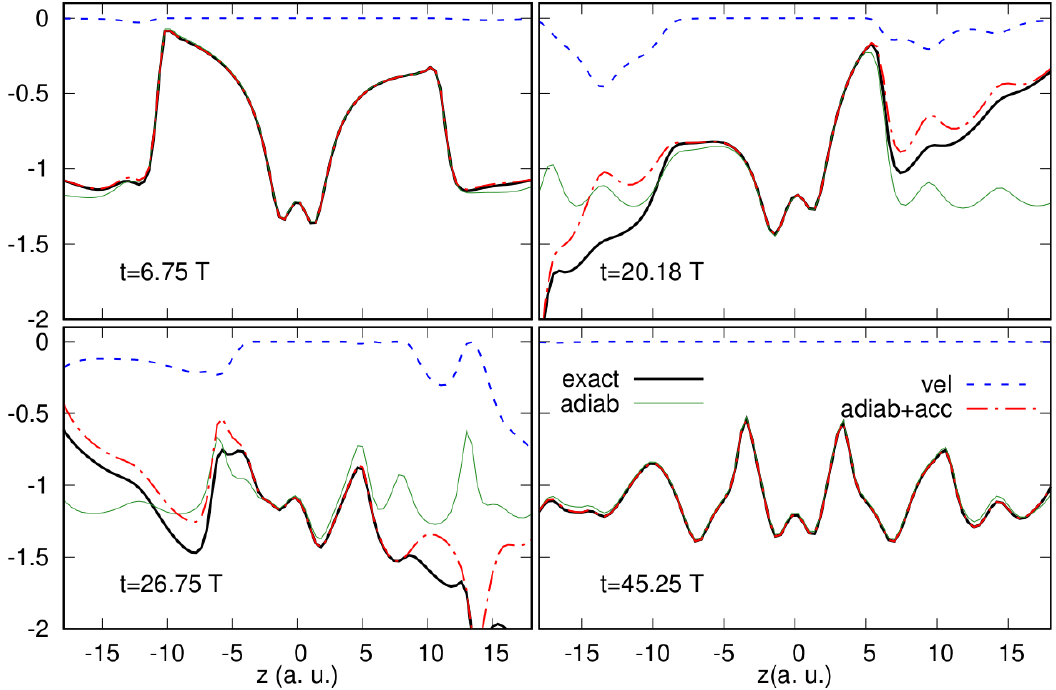}
\end{center}
\caption{The exact $e$-TDPES and its component from the marginal decompositions at various snapshots in time of  H$_2^+$.}
\label{fig:H2p-marginal}
\end{figure}

In Fig.~\ref{fig:H2p-marginal}, we present the terms/combinations-of-terms of marginal decomposition of the $e$-TDPES for the case of 
H$_2^+$ at four different snapshots in time.

{\bf Adiabatic term}: the adiabatic term (first term in Eq.~\ref{eq:exact_etdpes_tddft}) is the main constituent of the $e$-TDPES 
when it is decomposed according to Eq.~\ref{eq:exact_etdpes_tddft}. In this case, as it is seen in Fig.~\ref{fig:H2p-marginal} (upper left panel), 
it initially follows the exact $e$-TDPES in the inner-region and to some extent in the outer-region while it shows a different behavior in the 
asymptotic region (deep in the outer-region). However, as initially there is a negligibly small amount of electronic density far from the 
inner-region, this deviation does not influence the dynamics significantly. As the field intensity increases, the adiabatic potential starts to 
deviate from the exact potential, both in the inner-region and outer-region. It can be seen in Fig.~\ref{fig:H2p-marginal} 
(lower left panel) that the shape 
of the adiabatic potential in the inner-region (especially the up-field part)
and 
its (average) slope in the outer-region differ significantly from the exact $e$-TDPES. The (average) slope of the exact potential follows the slope 
of the field in the outer-region as expected. This feature together
with the up-field part the exact potential are mainly responsible for controlling the 
ionization from the up-field direction. 
Indeed the over-ionization corresponding to the propagation on the adiabatic potential discussed in the previous section
(see the lower left panel of Fig.~\ref{fig:sidf-comp-app}),  which starts around the $20$th 
optical cycle is associated to the lack of the asymptotic slope, and the error in the shape of the up-field well. In particular, the wrong asymptotic behavior of the
  adiabatic potential in the outer region, allows for ionization from
  both sides (up-field and down-field) in each half cycle, resulting in a huge overestimation of ionization probability.
However, an important feature of the exact $e$-TDPES 
is captured in the adiabatic potential: the development of the four wells representing the branching of the nuclear wavefunction 
in the inner-region of the potential which is associated to the correlation between the electronic and nuclear motions. Towards the end of the 
dynamics, where the field intensity is small again the adiabatic potential follows the exact $e$-TDPES closely as can be seen in Fig.~\ref{fig:H2p-marginal}
(lower-panel, right).

{\bf Velocity term}:  is the second term in Eq.~\ref{eq:exact_etdpes_tddft} and its overall contribution to the $e$-TDPES, in this case, 
is small especially in the inner region. As it can be seen in Fig.~\ref{fig:H2p-marginal}, it slightly corrects the adiabatic potential in the up-field (inner-)region 
as well as the outer-region but not enough to capture the essential features appearing in the exact $e$-TDPES.

{\bf Acceleration term}:  is the last term in Eq.~\ref{eq:exact_etdpes_tddft} that is initially very small in the inner-region 
but as the ionization sets in, the addition of this term to the adiabatic potential  significantly 
improves the shape of the potential, particularly when the field approaches the peak-intensity as evident in Fig.~\ref{fig:H2p-marginal} 
(upper-right  and lower-left panels)  in the inner-region as well as the outer-region (see the "adiab+acc''). The latter is due to the much better (average) asymptotic behavior of this term in the outer-region compared to the adiabatic term.
On the other hand, it also improves the up-field/down-field well in the inner-region when added to the adiabatic potential. As a result  propagating on the combination 
of adiabatic and acceleration terms  leads to  an ionization probability 
in a good agreement with the exact results as it is shown in Fig.~\ref{fig:sidf-comp-app} (lower left panel).

As the structures of the adiabatic, velocity, and acceleration terms in the case of A$_2^+$ and D$_2^+$ are very similar to H$_2^+$, we do not 
address them here.

 \subsubsection{\it x$_2^+$ case}
\begin{figure} 
\begin{center}
\includegraphics[width=0.5\textwidth]{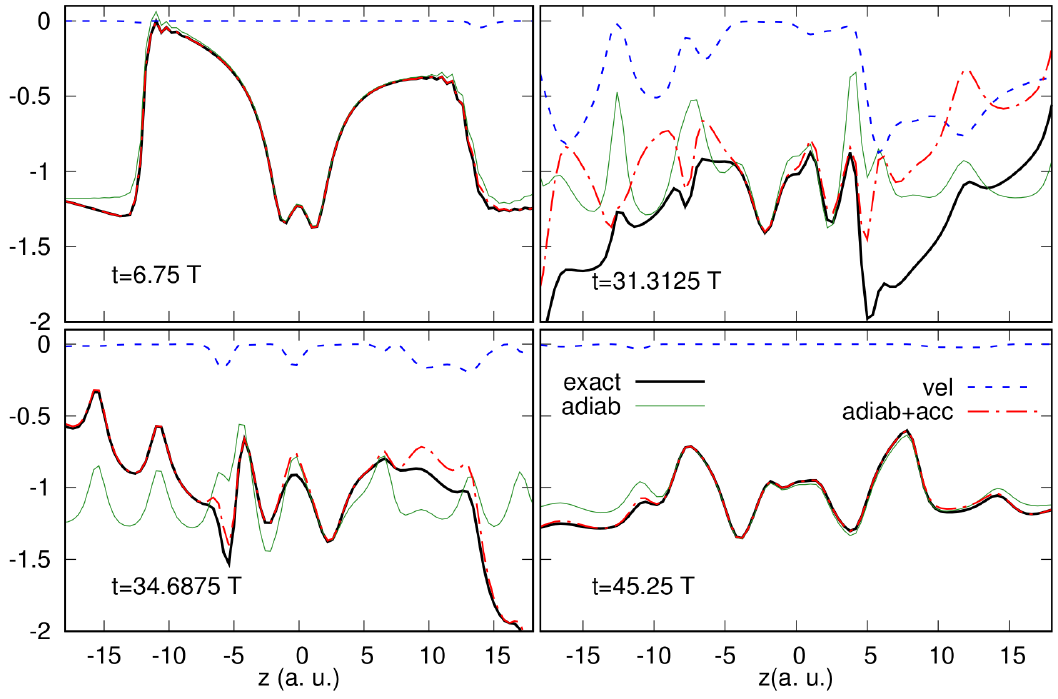}
\end{center}
\caption{ The exact $e$-TDPES and its component from the marginal decompositions at various snapshots in time of  x$_2^+$.}
\label{fig:c2p-marginal}
\end{figure}

In Fig.~\ref{fig:c2p-marginal}, the terms/combinations-of-terms of marginal decomposition of the $e$-TDPES for the case of 
x$_2^+$ are plotted at four different times. 

{\bf Adiabatic term}: behaves similarly to the case of H$_2^+$, i.e. initially and finally 
 it agrees well with the exact $e$-TDPES while when the field intensity is large it fails to follow the shape of the exact potential. Again, this 
 failure, in particular the wrong average slope of the adiabatic potential in the asymptotic region, results in a huge overestimation of the ionization probability.
   

{\bf Velocity term}: the velocity term plays a crucial role in this case as is seen in Fig.~\ref{fig:c2p-marginal} (lower-panel, left). 
In particular, as the field approaches its maximum intensity, it
exhibits more structure in the inner region and contributes more
significantly to the overall shape of the $e$-TDPES. Specifically, it
exhibits a valley in the center that is completely absent in the case
of H$_2^+$ and lowers the interatomic barrier when added to the adiab+acc
potential for most of the times between $t\approx28~T$ and
$t\approx40~T$ in which most of the ionization happens. The more dominant role of the velocity term in this case, could
  be attributed to the increase of ponderomotive (wiggling) motion of the electron driven by a stronger external laser field.

{\bf Acceleration term}: this term has a correct asymptotic behavior, similar to the case of H$_2^+$ but leads to a wrong estimate for the 
inner-barrier  (see Fig.~\ref{fig:c2p-marginal}, lower-left panel,) when added to the adiabatic term. The wrong estimation of the inner-barrier   
 is exaggerated after the field reaches its maximum intensity (especially due to the higher interatomic barrier around zero between $t\approx28~T$ and $t\approx40~T$) , 
leading to an inaccurate ionization probability after 
$t\approx30~T$, as seen in Fig.~\ref{fig:sidf-comp-app} (uper right panel).
 In ref.~\cite{SAMYG14}  the importance of the inner-barrier
(peaks) and up-field well (steps) to achieve 
the correct electron-localization has been shown. 

\section{Quasi-Classical Analysis}
\label{sec:quasi-cl}
The equations for the electronic wavefunction and conditional nuclear
wavefunction  cannot be solved exactly for systems of more
than a few degrees of freedom, just as solving the full molecular
Schr\"odinger equation exactly for those systems is not possible.  In many cases a
quasiclassical treatment of the nuclear dynamics should be sensible;
by quasiclassical, we mean an {\it ensemble} of classical
trajectories, weighted according to the initial distribution, is
evolved for the nuclei, rather than a single trajectory. This would
allow the possibility to capture spreading and branching of the
nuclear distribution. Such a procedure within the exact factorization
in its reverse flavor involves taking the classical limit of the conditional nuclear wavefunction which does not satisfy an equation of
Schr\"odinger form. Therefore, it differs from quasiclassical treatments of usual Schr\"odinger equations that have also been discussed in various forms for the marginal nuclear wavefunction within 
the "direct factorization" framework~\cite{AAG14,MAG15,AMAG16,SAMG15,AASMMG15,SW16}. Here we begin taking the classical limit of the conditional nuclear wavefunction by representing it in polar form: 
\ben
\chi_z(R,t) = A_z(R,t)e^{iS_z(R,t)/\hbar} 
\label{eq:sc_chiz}
\een
where $A_z(R,t)$ and $S_z(R,t)$ are both real functions. Inserting this into the equation of motion for $\chi_z(R,t)$, Eq.~\ref{eq:exact_ne}, and sorting the terms in orders of $\hbar$, we find to $O(\hbar^0)$:
\bea
\nonumber
&&\frac{1}{2\mu_n}\left(\frac{\partial S_z}{\partial R}\right)^2 + \frac{1}{2\mu_e}\left(\frac{\partial S_z}{\partial z}\right)^2 + V(z,R,t) -\epsilon_e(z,t) 
\\
&\,+& \frac{\tilde{p_e}(z,t)}{\mu_e}\frac{\partial S_z}{\partial z} + \left(\frac{\partial S_z}{\partial t}\right) = 0
\label{eq:Sz}
\eea
where we define $V(z,R,t) = W_{nn}(R) + W_{en}(z,R) +V_l(z,R,t)$, and $\tilde{p_e}(z,t) = -i\hbar \, \partial \, \Phi(z,t)/\partial z$ (see shortly for more on this term).  
Similarly, keeping only $O(\hbar^0)$ terms in Eq.~\ref{eq:exact_etdpes} for the $e$-TDPES, we find
\bea
\nonumber
\epsilon_e^{cl}(z,t) &=& \int dR \vert\chi_z(R,t)\vert^2\Big(\frac{1}{2\mu_n}\left(\frac{\partial S_z}{\partial R}\right)^2 + \frac{1}{2\mu_e}\left(\frac{\partial S_z}{\partial z}\right)^2 \\
&+& V(z,R,t) + \frac{\partial S_z}{\partial t}\Big).
\label{eq:sc_eps}
\eea
The terms on the right-hand-side correspond to classically evaluating $\mathcal{T}_n(z,t), \mathcal{K}_e^{\rm cond}(z,t), \epsilon^{\rm app}_e(z,t)$ and $\epsilon^{\rm gd}_e(z,t)$ respectively.  Notice that the electron-nuclear coupling operator $U_{en}^{coup}$ has a classical counterpart, as it contributes already at zeroth-order in $\hbar$ with the term $\frac{1}{2\mu_e}\left(\frac{\partial S_z}{\partial z}\right)^2$ in both Eq~(\ref{eq:Sz}) and~(\ref{eq:sc_eps}).

Eq.~\ref{eq:Sz} would be a  standard Hamilton-Jacobi equation of the
form $H({\bf q},\frac{\partial S_z}{\partial {\bf{q}}},t) +\partial
S_z/\partial t = 0$ (where $H({\bf q},{\bf p},t)$ is the Hamiltonian
function), for the action  $S_z(R,t)$ of two particles, one of mass
$\mu_n$ and the other of mass $\mu_e$ in a potential $V(z,R,t) -
\epsilon_e(z,t)$, if the second-last term on the left-hand-side was
not present. It is perhaps not surprising that we do not retrieve a
standard Hamilton-Jacobi equation, given that the equation for
$\chi_z$ is not a TDSE. Although an $\hbar$ multiplies this term, it does in fact contribute in the classical limit, as will be discussed shortly.
Still,  classical Newton-like equations can be derived from Eq.~\ref{eq:Sz} by defining the velocity fields, 
\ben
u_z^n(R,t) = \frac{1}{\mu_n}\frac{\partial S_z(R,t)}{\partial R}\;, {\rm and}\; u_z^e(R,t) = \frac{1}{\mu_e}\frac{\partial S_z(R,t)}{\partial z}
\label{eq:vel}
\een
Then, taking $\partial/\partial R$ of Eq.~\ref{eq:Sz} yields
\ben
\mu_n\frac{d u_z^n(R,t)}{dt}  = -\frac{\partial}{\partial R}\left(V(z,R,t)  + \tilde{p_e}(z,t) u_z^e(R,t)\right)
\label{eq:nuc_vel}
\een
and
\ben
\mu_e\frac{d u_z^e(R,t)}{dt}  = -\frac{\partial}{\partial z}\left(V(z,R,t) - \epsilon_e(z,t) + \tilde{p_e}(z,t) u_z^e(R,t)\right)
\label{eq:elec_vel}
\een
where $d/dt = \partial/\partial t + u_z^e\partial/\partial z + u_z^n \partial/\partial R$ is the time-derivative in the Lagrangian frame defined by the velocities.

\subsection{Quasiclassical analysis of the terms in the $e$-TDPES}
\label{sec:qs-anal-eTDPES}
Whether the equations above, together with the solution of the TDSE
Eq.~\ref{eq:eleq-ef} could form the basis of a  mixed
quantum-classical method remains for future work. Here, instead we
consider how a quasiclassical analysis of the entire coupled
electron-nuclear system can shed light on the structure and nuclear-mass-dependence of the terms
in the $e$-TDPES. Many aspects of electron dynamics in strong fields can be treated classically, especially when tunneling and quantization are not driving the primary physics.
Indeed, Ref.~\cite{VIC96} showed that classical trajectory
calculations reproduce the essential features of CREI for both cases
of fixed and moving nuclei.

To this end, we first evaluate $\tilde{p_e}(z,t)$ in Eq.~\ref{eq:Sz} to its lowest-order in $\hbar$. Inserting $\Phi(z,t) = a(z,t)\exp(i s(z,t))/\hbar$, analogous to Eq.~\ref{eq:sc_chiz}, into the TDSE Eq.~\ref{eq:eleq-ef} for the marginal wavefunction $\Phi(z,t)$, and keeping only the term that is lowest-order in $\hbar$, gives
\ben
\frac{1}{2\mu_e}\left(\frac{\partial s(z,t)}{\partial z}\right)^2 + \epsilon_e(z,t) + \frac{\partial s}{\partial t} =0 \;.
\label{eq:s}
\een
This is the Hamilton-Jacobi equation for the action $s(z,t)$ of a classical electron evolving in Hamiltonian $\tilde{T_e} +  \epsilon_e(z,t)$. Here $\tilde T_e = \frac{1}{2\mu_e}\left(\frac{\partial s}{\partial z}\right)^2$ is the electronic kinetic energy associated with the Hamiltonian in the TDSE for the electronic wavefunction, Eq.~\ref{eq:eleq-ef}. It is important to note that this is {\it not} the same as the true electronic kinetic energy (see Eq. 69 in Ref.~\cite{AMG12}). The term in Eq.~\ref{eq:Sz}, $-i\hbar\partial \Phi(z,t)/\partial z$ thus becomes $\tilde{p_e} \equiv \sqrt{2\mu_e\tilde{T_e}}$ in the classical limit. 

Next, we note that Eq.~\ref{eq:Sz} for $S_z$ is consistent with the corresponding classical limits taken for the electronic wavefunction and the full molecular wavefunction: Writing $\chi_z(R,t) = \Psi(z,R,t)/\Phi(z,t)$, with $\Psi(z,R,t) = A(z,R,t)\exp(iS(z,R,t))$ and $\Phi(z,t) = a(z,t)\exp(is(z,t))$, then 
\ben
S_z(R,t) = S(z,R,t) - s(z,t)
\label{eq:Sz_S_s}
\een
where, in the classical limit, $s(z,t)$ satisfies Eq.~\ref{eq:s} and $S(z,R,t)$ satisfies
\ben
\frac{1}{2\mu_n}\left(\frac{\partial S}{\partial R}\right)^2 + \frac{1}{2\mu_e}\left(\frac{\partial S}{\partial z}\right)^2 + V(z,R,t) + \left(\frac{\partial S}{\partial t}\right) = 0
\label{eq:S}
\een
It is straightforward to check by substitution that Eqs.~\ref{eq:s},\ref{eq:S},\ref{eq:Sz} and \ref{eq:Sz_S_s} are consistent with each other.  
Eqs.~(\ref{eq:s}) and (\ref{eq:S}) are both standard Hamilton-Jacobi equations, so, for the classical trajectories that they describe, we readily identify the following: 

\ben
\partial S/\partial R = P_n= \mu_n\dot{R},\nonumber
\een
the nuclear momentum; 
\ben
\partial S/\partial z = p_e,\nonumber
\een
the electronic momentum; 
\ben
\partial S/\partial t = E(t), \nonumber
\een
the classical energy; and, as in the earlier discussion, 
\ben
\partial s/\partial z = \tilde{p_e},\nonumber
\een
and 
\ben
\partial s/\partial t = \tilde{T_e} +\epsilon_e.\nonumber
\een
Now we imagine an ensemble of classical particles with initial
position and momenta distributed according to the initial wavefunction. We consider evaluating the various contributions to the
classical $e$-TDPES of Eq.~\ref{eq:sc_eps} from these classical
trajectories. The integral over $R$ in that equation then becomes a
sum over classical trajectories which have arrived at $z$ at time $t$,
but with differing $R$, i.e. one sums over all trajectories that have
reached $z$ at time $t$.

For the first term in Eq.~\ref{eq:sc_eps}, $\mathcal{T}_n(z,t)$, 
\begin{eqnarray}
\label{eq:sc_Tn}
\nonumber
\mathcal{T}_n^{cl}(z,t) &=& \int dR\vert\chi_z(R,t)\vert^2\frac{1}{2\mu_n}\left(\frac{\partial S_z}{\partial R}\right)^2 \nonumber \\
&=& \int dR\vert\chi_z(R,t)\vert^2 \frac{\mu_n}{2}\dot{R}^2 = 
\frac{\mu_n}{2 N^{traj}_z} \sum_{I_z}^{N^{traj}_z} \dot{R_{I_z}}^2 \nonumber \\
&&
\end{eqnarray}
where we noted that $\partial S_z(R,t)/\partial R = \partial S(z,R,t)/\partial R$ (from Eq.~\ref{eq:Sz_S_s}), and replaced the integral with a sum over all $N^{traj}_z$ trajectories $I_z$ that reach $z(t) = z$ at time $t$; $\dot{R_{I_z}}$ is the nuclear velocity along the $I_z$-th trajectory.
\begin{figure}
\begin{center}
\includegraphics[width=0.5\textwidth]{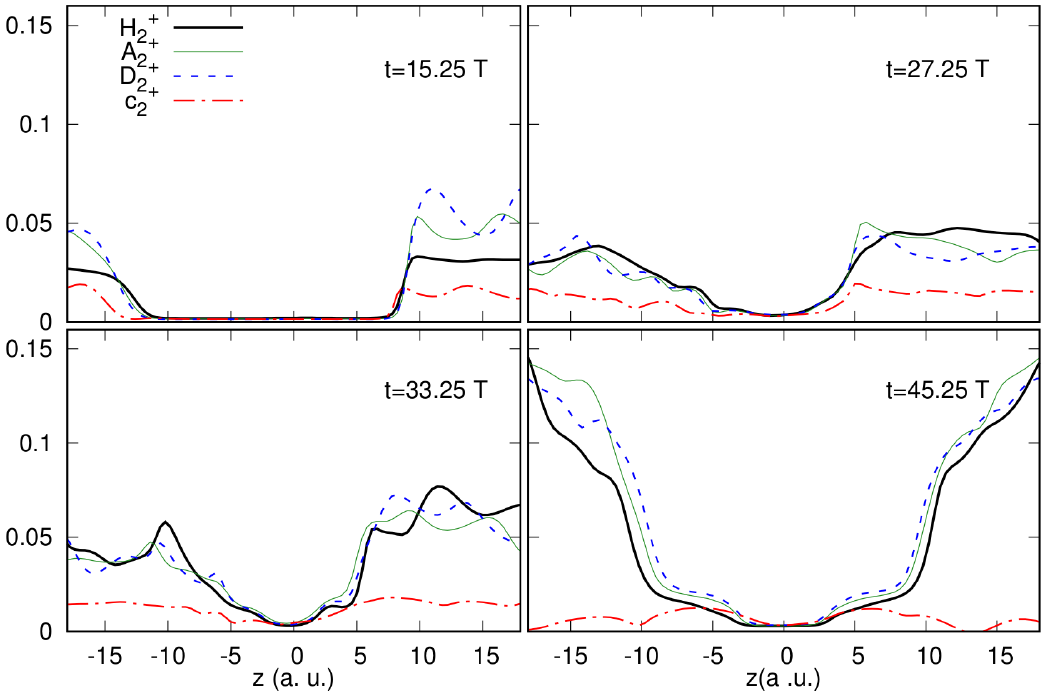}
\end{center}
\caption{ The kinetic term, $\mathcal{T}_n(z,t)$, at various snapshots in time for  H$_2^{+}$ and its isotopologues.}
\label{fig:Tn}
\end{figure}
First, let us see what this says about the overall structure of this
term.  During the dynamics, we observed that part  of the nuclear density oscillates around equilibrium, moving slower than the dissociating part of the density. This means that the trajectories for small $z$ are mostly associated with more slowly-moving nuclear dynamics near equilibrium, while those with larger $z$ are in the process of ionizing, 
and so are associated with faster nuclear speeds due to the net ionic charge they have left behind. Hence, for small $z$, there are more trajectories
with smaller $\dot{R}$'s than for large $z$, and so we expect a potential that rises as $z$ gets larger. This is indeed what we see in Fig.~\ref{fig:Tn}. 
Consider first the black curve, H$_2^+$. At early times, 
the middle region of $\mathcal{T}_n(z,t)$, where the bulk of the electron density is, is rather flat and takes its smallest value: Initially, the conditional nuclear probability has very little 
$z$-dependence in the region where the electron density is appreciable, i. e., at small times the classical positions of nuclear trajectories for $z$ in the region of appreciable electronic density are identical,  reflecting the broadness of the electron distribution relative to the nuclei. The small nuclear kinetic energy at the early times near equilibrium is essentially from zero-point motion. As the electrons begin to gain energy from the field and move out from the tails of the electronic distribution, 
the  $\mathcal{T}_n(z,t)$ grows accordingly since these tails are associated with trajectories where the nuclei are beginning to move apart under  Coulomb repulsion. The small middle flat region of 
$\mathcal{T}_n(z,t)$  gets narrower as ionization take away more of the electron density. As the final distribution sets in, $\mathcal{T}_n(z,t)$, one can identify three regions, reflecting the electronic 
distribution and the associated nuclear kinetic energies: an inner region associated with the density remaining near equilibrium, rising to the intermediate region where the shoulders of the electron density 
lie associated with the dissociated nuclear wavepacket, and then rising further out to the tails of the electron density which continue to oscillate in the field.

Regarding now the $\mu_n$-dependence: it might appear from Eq.~\ref{eq:sc_Tn} that $\mathcal{T}_n(z,t)$ grows as the mass $\mu_n$ increases, in the cases where the field is adjusted so that the average internuclear distance and speed remains the same. This is in fact not the case; in fact, the term remains about the same size, as the nuclear mass increases 
as is evident in Fig.~\ref{fig:Tn}. This is because, for larger $\mu_n$, the nuclear density tends to split less (see Fig.~\ref{fig:sidf-IR}), i.e. for larger $\mu_n$, a larger proportion of the nuclear density moves out to larger separations rather than  remaining near the origin with small $\dot{R}$. Yet about the same $\dot{\langle R \rangle}$ is maintained by design by the different field strengths, which means that the fastest trajectories in the distribution are slower for larger masses $\mu_n$  than for smaller $\mu_n$. 
This would contribute to a smaller rise of $\mathcal{T}_n(z,t)$ as one moves to larger $z$, as mass increases, compensating the larger $\mu_n$. 
In Fig.~\ref{fig:Tn} we plot $\mathcal{T}_n(z,t)$ for the different isotopologues at various snapshots in time. 
We see that the overall size of this term does not vary much with mass except for x$_2^{+}$. The exception is for the large mass $x_2^+$ where 
there is comparatively weak $z$-dependence of $\mathcal{T}_n(z,t)$. This is because the nuclear distribution moves largely as a whole, 
showing less difference between nuclear speeds at different parts of the distribution.

The corresponding analysis for the other three terms of Eq.~\ref{eq:sc_eps} is unfortunately not as straightforward, because the  evolution of the relevant classical trajectories themselves depend on the potential $\epsilon_e(z,t)$, which is the very object we wish to analyse! For example, the second term, $K_e^{\rm cond}(z,t)$, requires $\partial S_z/\partial z = (\sqrt{T_e} - \sqrt{\tilde{T_e}})$ but $\tilde{T_e}$ is the kinetic energy of an electron in the potential $\epsilon_e$. Future work along these lines will likely require further, possibly iterative, approximations to the potential, with the dual goal of analysing the structure of the terms and developing mixed quantum-classical schemes.

\section{Conclusions}
\label{sec:Conclusions}
The exact potential driving the electron dynamics within the exact factorization of the full electron-nuclear wavefunction
formally exactly accounts for the coupling to the nuclear subsystem as well as coupling to external fields. In order to develop
adequate approximations to treat electronic non-adiabatic processes, it is important to understand the structure of the exact potential and
pinpoint and analyse its important features in various situations. In this work, we have presented a detailed investigation of the
exact potential driving the electron dynamics in isotopologues of H$_2^+$ undergoing CREI and studied the dependence of
the correlated electron-nuclear dynamics on the nuclear mass.

The elaborate concept of time-resolved, $R$-resolved ionization probability,
$I(R,t)$, provides an extremely useful tool, indicating the
internuclear separations associated with ionization. The concept is analogous to the ionization probability calculated within the quasistatic picture
(with clamped nuclei) but can be used unambiguously for the fully
dynamical case. For the laser parameters and different isotopologues of H$_2^+$ discussed in this work, $I(R,t)$, 
exhibits only one peak. This is in agreement with most of the experimental findings~\cite{ERF05,BWSC08} and
in contrast to the predictions of CREI based on the standard clamped-nuclei quasistatic
picture. We suggest that $I(R,t)$ would be a very useful tool in future studies of CREI,
for example to resolve the laser parameters for which double-peak structure could
in fact be observed as in Ref.~\cite{XHK15}.

We found that for fixed laser parameters, the ionization yields 
rapidly decrease as a function of the mass of the isotopologue
because less of the nuclear density makes it to the CREI region during
the time the laser is on. For all the isotopologues, $I(R,t)$ indicated that the ionization is nevertheless dominated by the fraction of electrons associated with the nuclear
density in the CREI region defined  qualitatively by the original quasistatic argument. This implies that treating the nuclei as classical point particles will not work; one needs to account for their distribution, which, away from the large-mass limit, displays a branched structure with part of the distribution oscillating near equilibrium separation while part of it, associated with the CREI electrons, dissociates. 

Still, going beyond the quasistatic point nuclei picture and accounting for the nuclear distribution in an electrostatic way (as in Hartree-type approximations) is not adequate in capturing the dynamics accurately. The importance of going beyond an electrostatic description of electron-nuclear correlation is evident from our studies of
how different approximate electronic potentials perform in describing the CREI for
isotopologues of H$_2^+$. We have shown that, for the laser parameters used in
this work, one must go beyond any purely electrostatic treatment of electron-nuclear correlation, and include truly dynamical 
aspects of the nuclear distribution and its coupling to the electronic system to get a good prediction of the ionization. In determining errors from conventional approximations and deviations of approximate potentials from the exact potential, what are more important than the nuclear-to-electronic mass ratios, are the nuclear velocities in the wavepacket.

There are many different approaches to developing approximate methods based on the exact factorization. For example, one may consider the relative importance of different components of the exact potential when decomposed in terms of the conditional wavefunction as in Ref.~\cite{KAM15} (there, it was found that generally all terms were important). One may consider also approximations based on the marginal decomposition presented here, where we found that the "adiab+acc" component of the marginal decomposition describes the
electronic dynamics accurately in all cases we studied here with the exception of a fictitious isotopologue with an effective nuclear mass 10 times larger than H$_2^+$. Further
investigation of this potential may lead to development of an adequate approximation for practical purposes
%
capable of describing the ionization dynamics accurately. 
Another approach is to develop quasiclassical approximations, and here
we have sketched out a semiclassical derivation of the electronic and conditional nuclear equations of the exact factorization in its reverse
form and analysed the structure of the nuclear kinetic term of the exact potential semiclassically.  

This work has highlighted the effect of the complex interplay of electronic and nuclear dynamics in strong field 
enhanced ionization processes by demonstrating the large differences in the conventional potentials and the dynamics 
they cause with the exact potential driving the electron for systems of varying nuclear mass. The explorations of the 
details of the potential and approximation methods lay the ground-work for future development of accurate methods 
for coupled electron-ion dynamics in non-perturbative fields. Whether  the dynamical electron-nuclear effects play
as crucial a role for polyatomic molecules, and the scaling of these effects with respect to the number of electrons
 and number of nuclei, is also an important avenue for future research 

\acknowledgments{We acknowledge support from the European Research Council(ERC-2015-AdG-694097),  Grupos Consolidados (IT578-13), 
and the European Union's Horizon 2020 Research and Innovation programme under grant agreement no. 676580. A.K. and A.A acknowledge funding from the European Uninos Horizon 2020
research and innovation programme under the Marie Sklodowska-Curie grant agreement no. 704218 and 702406, respectively. N.T.M thanks the National Science Foundation, grant CHE-1566197 for support.}

\bibliography{./ref}

\end{document}